\documentclass[10pt,leqno]{amsart}
\usepackage{graphicx}
\usepackage{indentfirst,csquotes}

\usepackage{amssymb,amsthm,amsmath}
\usepackage{xcolor,paralist,hyperref,titlesec,fancyhdr,etoolbox}
\usepackage{booktabs,caption,tabularx}  

\titleformat{\section}[display]{\normalfont\huge\bfseries\centering}{\centering\chaptertitlename\thechapter}{10pt}{\Large}
\titlespacing*{\section}{0pt}{0ex}{0ex}

\hypersetup{ colorlinks=true, linkcolor=black, filecolor=black, urlcolor=black }

\usepackage{lipsum}
\begin{document}
\title{Determination of geopotential by hydrogen masers based on global navigation satellite system} 
\maketitle

Lei Wang\textsuperscript{1},
Wei Xu\textsuperscript{1,2},
Li-Hong Li\textsuperscript{1},
Peng-Fei Zhang\textsuperscript{1},
An Ning\textsuperscript{1},
Rui Xu\textsuperscript{1},
Peng Cheng\textsuperscript{1}
Zi-Yu Shen\textsuperscript{3},
Wen-Bin Shen\textsuperscript{1,4,*}\\
\it{1 } Time and frequency Geodesy Center, School of Geodesy and Geomatics, Wuhan University, Wuhan 430079, China.
\\
\it{2 } School of Geographic Information and Tourism, Chuzhou University, Chuzhou 239000, China.
\\
\it{3 } School of Resource, Environmental Science and Engineering, Hubei University of Science and Technology, Xianning 437100, China.
\\
\it{4 } State Key Laboratory of Information Engineering in Surveying, Mapping and Remote Sensing, Wuhan University, Wuhan 430079, China.
\\
\bf{E-mail: wbshen@sgg.whu.edu.cn} 

\rm

\begin{abstract}
According to the general relativity theory, a clock runs faster at a position with higher geopotential. Thus, the geopotential difference can be determined by comparing the clock’s vibration frequencies. Here we report on the experiments to determine the geopotential difference between two remote sites based on precise point positioning time-frequency transfer technique. The experiments include the remote clock comparison and the local clock comparison using two CH1-95 active hydrogen masers acted as the reference clocks of global navigation satellite system time-frequency receivers. Considering the local clock comparison as calibration, the determined geopotential difference between two sites by our experiments is 12,142.3 (112.4) m$^{2}$/s$^{2}$, quite close to the value 12,153.3 (2.3) m$^{2}$/s$^{2}$ computed by the EIGEN-6C4 model. Results show that the proposed approach here for determining geopotential difference is feasible, operable, and promising.
\end{abstract} 


\bigskip

\section*{Introduction}

The general relativity theory (GRT)\cite{1} states that a clock at a position with higher geopotential ticks quicker than the identical clock at a position with lower geopotential\cite{2}. Previous studies have shown that one can use the gravity frequency shift approach\cite{3} to determine the geopotential difference between arbitrary two sites where precise clocks are set by comparing the clock’s vibration frequencies\cite{4,5,6,7}. This approach provides advantages in some aspects to overcome the drawbacks of the conventional approach that needs combining leveling with additional gravimetry\cite{8}, which is limited by the error accumulation with the increase of the length of the measurement-line and the impediment of mountains, rivers, and oceans when two points are located at separated islands or continents\cite{3}. However, actual application of the new approach requires high-performance clocks with a fractional frequency accuracy of 1×10$^{-17}$ to 1×10$^{-18}$ at least, corresponding to a resolution of about 1 m$^2$/s$^2$ to 0.1 m$^2$/s$^2$ in geopotential\cite{7}. 

With quick development of science and technology of atomic clocks, the frequency instability for state-of-the-art optical clocks has reached 10$^{-19}$ level\cite{9,10,11}, and the systematic uncertainties has achieved 10$^{-18}$  and below 10$^{-18}$ level\cite{10,12,13,14,15}, corresponding to an error equivalent to less than 1 s over the lifetime of the Universe\cite{16}. Several groups in the world completed relevant experiments to determine the geopotential difference using optical clocks based on the optical fibre frequency transfer (OFFT) approach\cite{4,5,7}. These experiments have demonstrated that the results by adopting the OFFT approach are consistent well with the independent measurements by the conventional approach, which indicates that OFFT is prospective and could be practically applied to determining the geopotential. By OFFT approach, it is necessary to compare two clocks located at separated sites connected with optical fibres. If the two sites are far away or located at different continents separated by oceans, it is inconvenient or difficult to connect clocks with optical fibres. Additionally, it is almost unrealistic to construct an optical fibre network to connect two arbitrary sites due to the high expense, especially anywhere in a country with a vast territory. Therefore, the satellite-based time-frequency transfer plays significant role\cite{17,18,19}, including global navigation satellite system (GNSS), communication satellites, and other satellite constellations that thanks to the fast development of space science and technology. 

In recent decades, the GNSS techniques have been applied extensively to high-accuracy time-frequency transfer since the International GNSS Service (IGS) has provided precise GNSS orbit and clock products\cite{20}. With the rapid development of the precise point positioning (PPP) technique in recent years\cite{21,22,23}, comparing with the PPP technique solving the carrier phase ambiguities as floating numbers that limits the performance of the time-frequency transfer to the low 10$^{-16}$, the PPP ambiguity resolution (PPP-AR) technique has been proposed to recover the integer nature of the carrier phase ambiguities\cite{24}. According to the latest results, the PPP-AR technique can provide time-frequency transfer instability with an improved long-term performance of order 7×10$^{-16}$/$T$ for clock comparison, where $T$ is the duration in days of continuous phase measurements, thus reaching a sub 10$^{-16}$ level after one week of averaging\cite{25} and can be applied for determination of geopotential. Several different kinds of experiments related to satellite-based time-frequency transfer techniques have been carried out to determine geopotential difference, including for instance, GNSS common-view (CV)\cite{26}, two-way satellite time and frequency transfer (TWSTFT)\cite{27}. Here, we report new experiments for determining the geopotential difference between two sites using two hydrogen masers based on the PPP time-frequency transfer (PTFT) using PPP-AR technique.

\bigskip

\section*{Experiments}

The experiments are carried out between Jiugongshan Time-Frequency Station (JTFS) in Xianning and Luojiashan Time-Frequency Station (LTFS) in Wuhan, China. The straight-line distance between the two stations is approximately 129 km, and the height difference between them is about 1,245 m. We employed two parallel independent sets of instruments, denoted as equipment set I and equipment set II. Both equipment sets have the same specifications and performances to avoid or weaken the system errors induced by the instruments. Each equipment set includes a CH1-95 active hydrogen maser (AHM) (C1/C2), a GNSS time-frequency receiver, a choke-ring antenna and relevant cables (FIGURE. 1). The hydrogen masers were manufactured by the factory KVARZ in Russia, and the production (nominal) frequency instability of either hydrogen maser is less than 5×10$^{-16}$ @ day. After being transported from KVARZ to Beijing in China, the performances of the two hydrogen masers, officially inspected by the National Institute of Metrology, China, are consistent with the nominal ones (FIGURE. 2e). The consistency demonstrates that the transportation process has very little or even no effect on the performances of the hydrogen masers, which provides strong supports for the experiments. The hydrogen masers provided time and frequency standards with one pulse per second (1PPS) and 10 MHz for GNSS time-frequency receivers, and the receivers can obtain the pseudorange and carrier phase observations through the choke-ring antennas that were applied to reducing the multipath effects while receiving navigation signals from visible GNSS satellites. These equipment sets were installed at JTFS or/and LTFS, respectively, to form the GNSS observation stations for time-frequency transfer. For convenience, when equipment set N (N=1, 2) was installed at JTFS/LTFS, the GNSS station is denoted as J00N/L00N.

We implemented both remote clock comparison (RCC) and local clock comparison (LCC) denoted as AB experiment and AA experiment, respectively. The schematic outline of experiments is shown in FIGURE. 1. In both AB and AA experiments, the CH1-95 AHM and GNSS time-frequency receiver worked in well-conditioned indoor with relative humidity of 60$\%$ and stable temperature of about 25 °C, the main environment requirements for the CH1-95 AHM working stably, to largely keep from potential environment disturbances, and the choke-ring antenna was set on the roof for good observation sight.
\begin{figure}
\centering
\includegraphics[width=.8\linewidth]{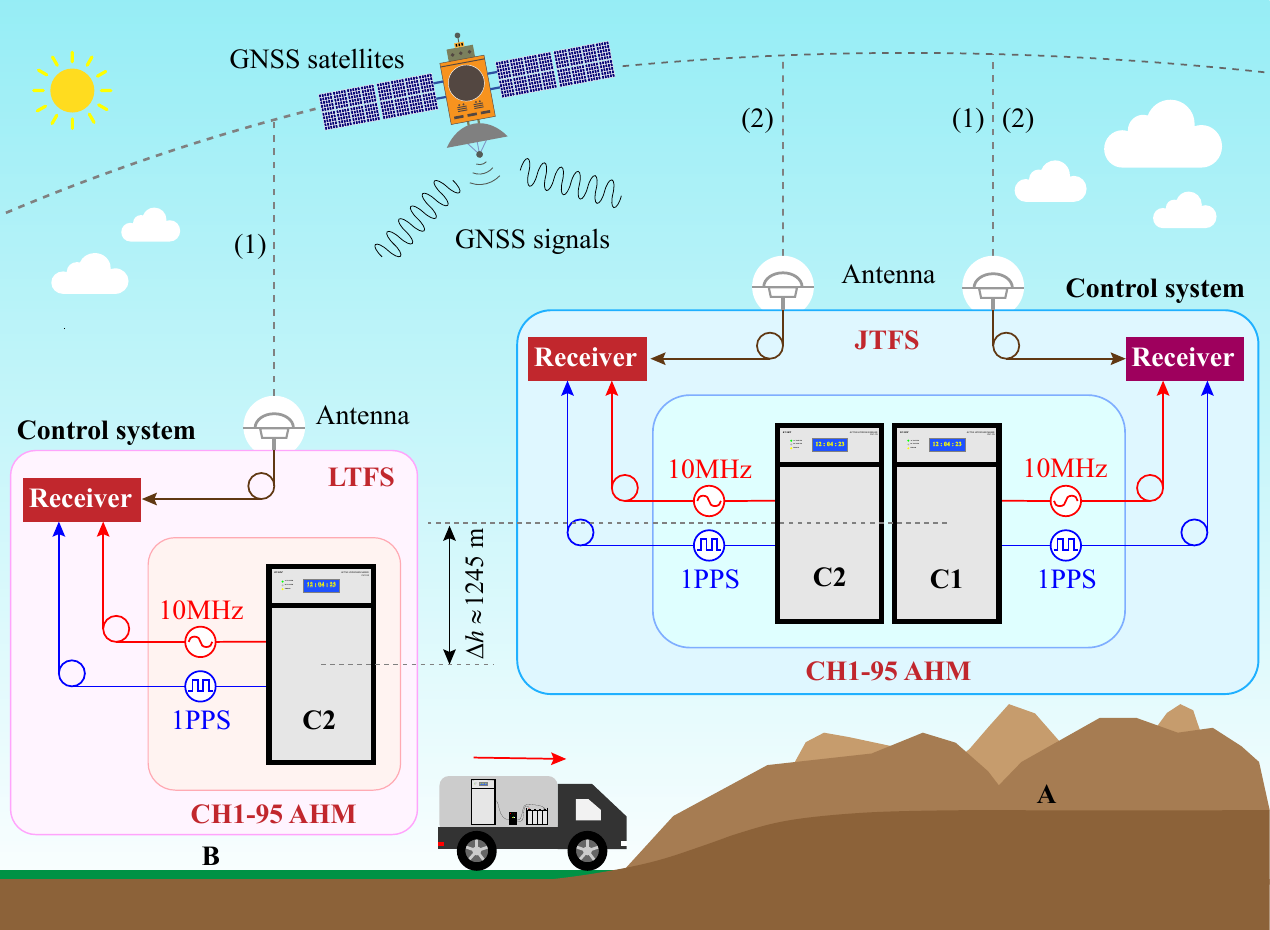}
\caption{Schematic of the experimental setups and process. Two sets of instruments, each including a CH1-95 AHM, denoted as C1 or C2, a GNSS time-frequency receiver, a choke-ring antenna and relevant cables to form a GNSS observation station for time-frequency transfer. Control systems are used to avoid potential environmental influences. (1) In AB experiment, C1 and C2 are used as external clocks for GNSS stations J001 at JTFS and L002 at LTFS, respectively. Clock comparison is conducted by PTFT using GNSS pseudorange and carrier phase observations. (2) In AA experiment, the whole set of instruments at LTFS is transported to JTFS with an air-conditioned vehicle (along the red arrow). C2 is placed next to C1 and forms a new GNSS station J002 at JTFS.}
\end{figure}

In AB experiment, the equipment set I was installed at JTFS and the equipment set II was installed at LTFS to construct GNSS stations J001 and L002 to form the remote time-frequency transfer link, and the observations lasted for 14 days. For calibration, we need AA experiment to determine the inherent frequency offset between C1 and C2 without any adjustments. The equipment set II was transported to JTFS from LTFS using an air-conditioned vehicle for about four-hours journey, during which the C2 always ran normally powered by an uninterruptible power supply (UPS). Then, the equipment set II was installed at JTFS, where C2 was placed next to and at the same height as C1 (FIGURE. 1). A new GNSS station J002 at JTFS was constructed and formed the local time-frequency transfer link with J001 that had been constructed, and the observations also lasted for 14 days in AA experiment. The two hydrogen masers ran continuously and freely without any artificial adjustments throughout AB and AA experiments, which is a critical step to determine geopotential difference using atomic clocks. 

\bigskip

\section*{Data Analysis}

In AA experiment, at beginning the C2 ran less reliably due to the vibration and shaking during the four-hours transportation, and about two days later it ran with good performance. Here, we selected continuous and stable observations data with quality control (enough visible satellites, less signal disturbance, etc., see Supplementary Information (SI)) to resolve the clock offset series of GNSS stations, and then determine the time difference series between two clocks (see Methods). In AB and AA experiments, we selected the GPS data segments from days of year (DOY) 59 to 69 and 76 to 85 in 2021, respectively. The sampling interval of all GPS observations used in this study is 30 s. Previous studies and experiments have shown that more visible satellites and smaller time dilution of precision (TDOP) values can improve the accuracy of the GNSS positioning and timing and thus improve the frequency stability of time-frequency transfer link between two stations\cite{28} . Each GNSS station in AB or AA experiment can observe more than 7 GPS satellites every day, and the TDOP values of all stations are less than 1.7 (see SI). These demonstrate that all GNSS stations have excellent observation conditions.

\begin{figure}
\centering
\includegraphics[width=.8\linewidth]{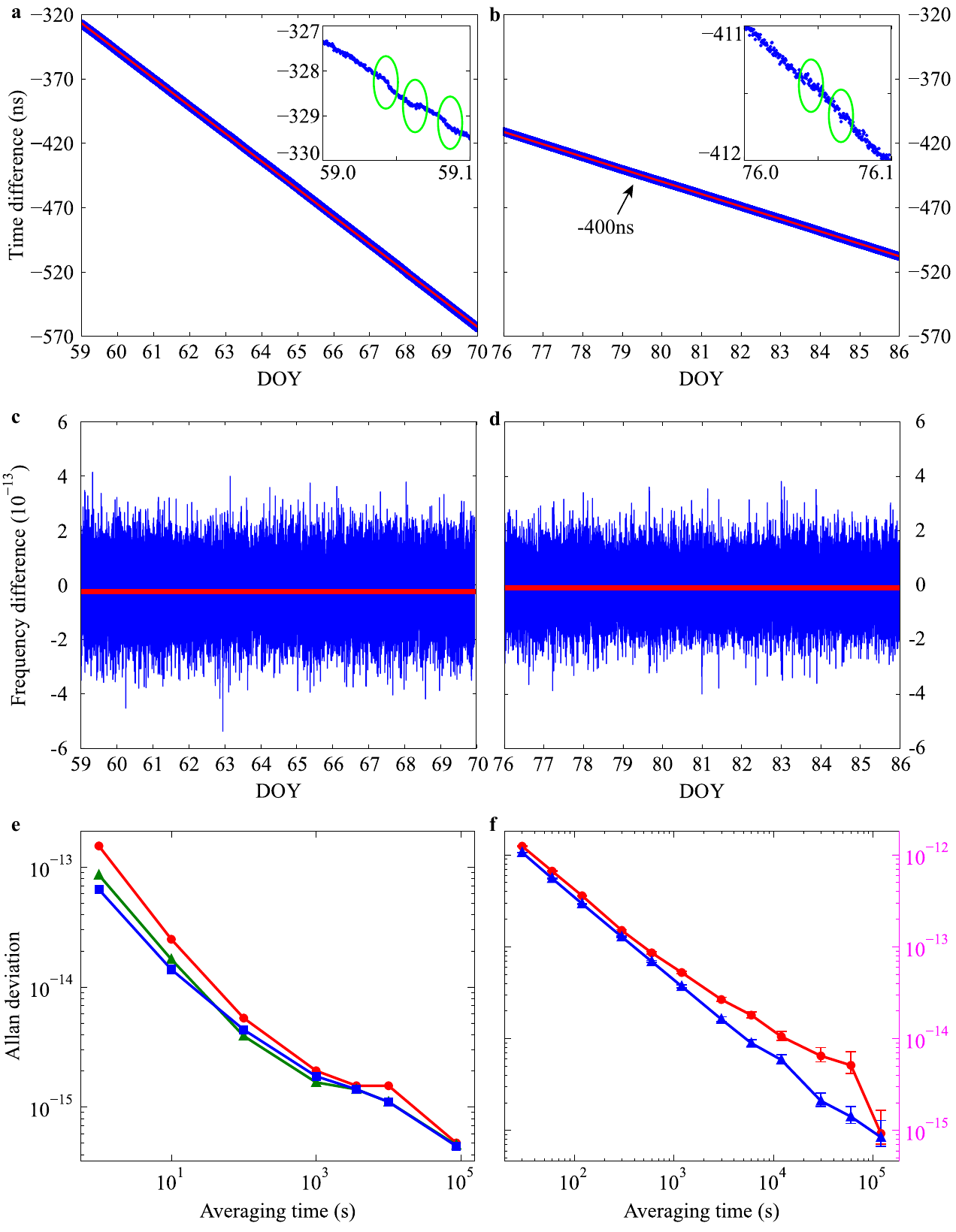}
\caption{Time and frequency difference series and frequency instability of clock comparisons. a, Time difference series (blue dots) and the fitting line (red line) in AB experiment from DOY 59 to 69 in 2021. Inset shows the time difference series of 0.1 day in DOY 59, 2021 and green circles mark jumps. b, Time difference series (blue dots) and the fitting line (red line) in AA experiment, whose y-axis subtracting 400 ns shares the same y-axis with a. Inset shows the time difference series of 0.1 day in DOY 76, 2021 and green circles mark jumps. c, Frequency difference series (blue curve) and the average value of frequency differences (red line) of the link in AB experiment where the frequency drift has been removed. d, Frequency difference series (blue curve) and the average value of frequency differences (red line) of the link in AA experiment where the frequency drift has been removed. e, Frequency instability of the CH1-95 AHM. The nominal frequency instability of either hydrogen maser (red curve with dots) and the frequency instabilities inspected authoritatively (green curve with triangles and blue curve with squares) expressed as Allan deviation (ADEV) present excellent consistency. f, Frequency instability of links in AB (red curve with dots) and AA (blue curve with triangles) experiments expressed as ADEV. The error bars show the   uncertainty.}
\end{figure}

Based on the PTFT, the time difference between the hydrogen maser (as the external clock for receiver) at the station of interest and the reference time of precise products used in resolving is called receiver clock offset in PPP, and it can be estimated as a parameter every epoch to form the clock offset series (see Methods). Then the time difference series between two hydrogen masers can be obtained by differencing the clock offset series of two stations (see Methods) and consequently the time difference series can be transformed to the frequency difference series (see SI). To determine the frequency shift between C1 and C2 with high accuracy and stability, we applied the latest open-source software, PRIDE PPP-AR\cite{29}, to processing the GPS observations (see Methods) with excellent ambiguity resolution (see SI). Then, the time difference series between C1 and C2 in AB and AA experiments were obtained (FIGURE. 2a-b), from which we determined the change rate of time difference after removing the quadratic term, equivalent to the frequency drift between C1 and C2, by directly linear fitting with least-squares (see Methods). The gravity frequency shift caused by geopotential difference between JTFS and LTFS comes from the frequency shift in AB experiment subtracting the frequency offset in AA experiment. For this resolution, some jumps occurred in time difference series (such as the insets in FIGURE. 2a-b) can influence the accuracy of the determined frequency shift. Hence we take the following measures: 1) transform time difference series to frequency difference series (see SI); 2) replace frequency difference jump with linearly interpolated values determined by the two values before and after any interior jumps; 3) remove the frequency drift; 4) compute the average value of frequency differences (FIGURE. 2c-d). Thus, the average values of frequency differences demonstrate respectively the frequency shift in AB experiment and the frequency offset in AA experiment between C1 and C2, and the gravity frequency shift between A and B can be obtained by removing the frequency offset in AA experiment from the frequency shift in AB experiment. Based on the determined gravity frequency shift, we may determine the geopotential difference between JTFS and LTFS.

\bigskip

\section*{Result}

As shown in FIGURE. 2a-b, the time differences between C1 and C2 in AB and AA experiments are reduced by about 235 ns for 11 days and about 97 ns for 10 days, respectively. The AA experiment demonstrates that there exists the inherent frequency offset between C1 and C2, and therefore AA experiment is meaningful and necessary as calibration for determining the geopotential difference. As shown in FIGURE. 2e-f, the frequency instabilities of links in AB and AA experiments expressed as ADEVs computed by Stable32 software (http://www.stable32.com/) can respectively reach 9.2468 ×10$^{-16}$ and 8.4241 ×10$^{-16}$ at the averaging time of 120000 s, which are roughly consistent with the nominal frequency instability of the hydrogen masers (FIGURE. 2e). These results demonstrate that the PTFT has taken full advantages of the performances of CH1-95 AHM. The resolved frequency differences between C1 and C2 in AB and AA experiments (FIGURE. 2c-d), derived from the average values of frequency differences transformed from time difference series, are -2.4729×10$^{-13}$ and -1.1219×10$^{-13}$, respectively. Thus, the gravity frequency shift caused by the geopotential difference is -1.3510×10$^{-13}$ with an uncertainty of 1.2501×10$^{-15}$ derived from the ADEVs of links in AB and AA experiments based on the error propagation law. Hence, the determined geopotential difference between A and B is 12142.3 (112.4) m$^2$/s$^2$ (the numbers in the parentheses are the 1 $\sigma$ uncertainties referred to the corresponding last digits of the quoted results\cite{5}). Compared with the corresponding model value 12153.3 (2.3) m$^2$/s$^2$ computed by the EIGEN-6C4 model\cite{30} (see Methods), the deviation is 11.0 (112.4) m$^2$/s$^2$ (equivalent approximately to 1.1 (11.5) m in height).

\bigskip

\section*{Conclusion}
Our experiments have shown an approach to directly determine the geopotential difference between two remote sites based on the PTFT using active hydrogen masers. Although the accuracy at present is not high enough compared with the conventional approach, the experiments here have demonstrated a feasible, convenient, and economic approach to determine geopotential difference between arbitrary two points. The accuracy of the results is mainly influenced by the performances of the hydrogen masers, the observation noises induced by GNSS satellite signals, observation environments and instruments. Thanks to quick development of time and frequency science and technology in recent years, high-performance clocks (including portable optical clocks) with frequency instability at 10$^{-19}$ level have been successively generated\cite{9,10,11}. The free-space dissemination of time and frequency with 10$^{-19}$ instability over 113 km has been reported that could lay the groundwork for satellite time–frequency dissemination\cite{31} and the fractional frequency measurement uncertainty of 7.6 × 10$^{-21}$ has been reached\cite{16}. Hence, in the future the conventional approach could be supplemented or even replaced by PTFT using high-performance clocks and advanced techniques, greatly impacting future the way of determining geopotential and promoting the developments of time-frequency science and geodesy.

\bigskip
\bigskip
\section*{Methods}
\bigskip

\subsection*{Geopotential difference determination based on gravity frequency shift.}
Suppose two atomic clocks,   and  , have been a priori calibrated, and they are placed at two points 1 and 2, respectively. After a standard time duration  , accurate to the level of  , the geopotential difference between these two clocks can be expressed as\cite{2,3}:
\begin{equation}
\Delta W_{12}=\left(W_2-W_1\right)\approx-\frac{\Delta t_{12}}{T}c^2\approx-\frac{\Delta f_{12}}{f_0}c^2
\end{equation}
where $W_1$ and $W_2 $ denote the geopotentials at points 1 and 2, respectively; $\Delta t_{12}=t_2-t_1$ is the time difference between C1 and C2 , where $t_1$ and $t_2$ are respectively the time durations recorded by clocks C1 and C2 after a time duration $T$; $\frac{\Delta f_{12}}{f_0}$ is the (relative) gravity frequency shift between C1 and C2, where $\Delta f_{12}=f_2-f_1$ and $f_0$ is the nominal frequency of the atomic clock; $c$ is the speed of light in the vacuum.

\bigskip

\subsection*{Geopotential difference determination by global gravity field model EIGEN-6C4.}
The geodetic coordinates of the GNSS station $i\left(i=1,2\right)$ can be resolved along with clock offset while estimating parameters. However, that is the coordinates at the center of GNSS receiver antenna, where the geopotential $W_{ant}$ and gravity $g_{ant}$ at antenna can be computed online applying global gravity field model EIGEN-6C4 from ICGEM website (https://icgem.gfz-potsdam.de/). The vertical height difference between GNSS receiver antenna and the maser cavity of CH1-95 AHM, $\Delta H=H_{ant}-H_m$, is measured by laser range finder, where $H_{ant}$ and $H_m$ are the vertical heights above the geoid of antenna and maser cavity, respectively. Then, the geopotential at the maser cavity of CH1-95 AHM C1/C2 can be obtained as $W^i_m\approx W^i_{ant}+g^i_{ant}\cdot\Delta H^i$, and the geopotential difference between hydrogen masers is $\Delta W^{12}_m= W^2_m-W^1_m$. 

\bigskip

\subsection*{Precise point positioning time-frequency transfer (PTFT).}
As the method of determining geopotential difference based on GRT described, the key is to compare two atomic clock’s vibration frequencies. The PTFT can be applied to completing this task. With this technique, the clock offset between the atomic clock time $\left(t_A\right)$ and the reference time of precise products $\left(t_r\right)$ can be estimated as a parameter using GNSS pseudorange and carrier phase observations\cite{22,32}. The atomic clock can provide time and frequency standards for the GNSS time-frequency receiver, and $t_r$ is a very stable ensemble time scale released by data analysis center, here used is IGS data canter of Wuhan university (http://www.igs.gnsswhu.cn/). The clock offset between $t_A$ and $t_r$ can be expressed as $t^P_A-t_r$ for any GNSS station, where $t^P_A$ denotes the local time recorded by the atomic clock at point $P$. Thus, the time difference $\Delta t^P_Q$between two atomic clocks at any two points $P$ and $Q$ can be computed by differencing $t^P_A-t_r$ and $t^Q_A-t_r$, expressed as\cite{32}:
\begin{equation}
\Delta t_{PQ}=t^Q_A-t^P_A=\left(t^Q_A-t_r\right)-\left(t^P_A-t_r\right)
\end{equation}
After a standard time duration $T$, the change rate of time difference, $\Delta \frac {t_{PQ}} {T}$, can be determined by linear fitting with least square. According to equation (2), $t_r$ is used only as a reference or “bridge”. Therefore, arbitrary two clocks can be compared based on the PTFT technique, no matter where the clocks are located. In addition, the PTFT is easy to put into operation that is not required for synchronous observations and can provide excellent short-term and medium-term frequency stability. Therefore, this technique can be prospectively adopted and will be widely applied for determining geopotential difference in the future.

\bigskip

\section*{Supplementary information}

\bigskip

\subsection*{Visibility of GPS satellites.}
FIGURE. S1 presents the daily average values of the number of visible GPS satellites and TDOP values of GNSS stations in AB and AA experiments. The visibility of the GPS satellites at station L002 in AB experiment is worse than that at station J002 in AA experiment. Meanwhile, the TDOP values at L002 are larger than that at J002, which means that the time-frequency transfer accuracy of L002 is worse than J002. The visibility of the GPS satellites and TDOP values at J001 and J002 are nearly the same, respectively. Hence, the frequency stability of the link in AA experiment is better than that in AB experiment, which is demonstrated in FIGURE. 2f.   

\renewcommand\thefigure{S\arabic{figure}}  
\setcounter{figure}{0}
\begin{figure}[tbhp]
\centering
\includegraphics[width=.8\linewidth]{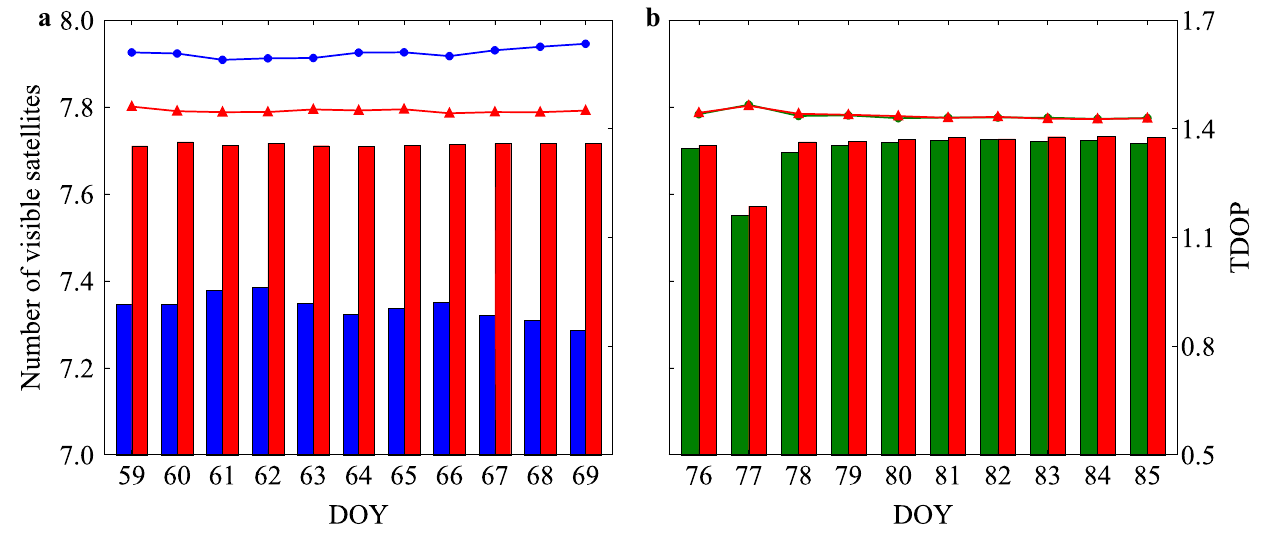}
\caption{Daily average values of the number of visible satellites and TDOP values of GNSS stations. a, Number of visible GPS satellites (histograms) and TDOP values (curves with points) of GNSS stations J001 (red) at JTFS and L002 (blue) at LTFS in AB experiment from DOY 59 to 69 in 2021. b, Number of visible GPS satellites (histograms) and TDOP values (curves with points) of GNSS stations J001 (red) and J002 (green) at JTFS in AA experiment from DOY 76 to 85 in 2021.}
\end{figure}


\subsection*{Precise point positioning ambiguity resolution (PPP-AR) technique.}
In our experiments, to process the GPS pseudorange and carrier phase observations reliably and accurately using the PPP-AR technique, we applied the latest open-source software, PRIDE PPP-AR\cite{29}. This software has been developed by the PRIDE Lab at the GNSS Research Center of Wuhan University. It can provide float solutions with wide-lane and narrow-lane ambiguity estimates and integer ambiguity resolution recovering the integer nature of single-station ambiguities by using the phase clock/bias products. PRIDE PPP-AR has been demonstrated that the ambiguity-fixed positions outperform those by float solutions. According to the PPP parameters estimation, the accuracy of the ambiguity-fixed clock offset can also be improved and performed well in time-frequency transfer links. The strategies and options of GPS observation data processing using PRIDE PPP-AR are presented in TABLE S1.  

\renewcommand\thetable{S\arabic{table}}  
\setcounter{table}{0}
\begin{table*}[tbhp]
\centering
\caption{Summary of the GPS observation data processing strategies and options for PRIDE PPP-AR}
\begin{tabular}{l p{7cm}}
\midrule
Items & Strategies or options \\
\midrule
Solutions & PPP-AR \\
Observables & {Ionosphere-free combination of GPS pseudorange and carrier phase\centering} \\
Cutoff elevation & 7° \\
Sampling interval & 30 seconds \\
Strict editing & YES \\
Zenith tropospheric delay & PWC: 60 minutes \\
Horizontal tropospheric gradients & PWC: 720 minutes \\
Troposphere mapping function & Vienna Mapping Function 3 (VMF3) \cite{33}\\
Correcting second-order ionospheric delays & YES \\
Ambiguity fixing & ROUNDING \\
Ambiguity duration & 600 seconds \\
Cutoff elevation for PPP-AR & 15° \\
Widelane decision & Deviation: 0.2; Sigma: 0.15 (Cycle) \\
Narrowlane decision & Deviation: 0.15; Sigma: 0.15 (Cycle) \\
\bottomrule
\end{tabular}
\end{table*}

\begin{figure}
\centering
\includegraphics[width=.8\linewidth]{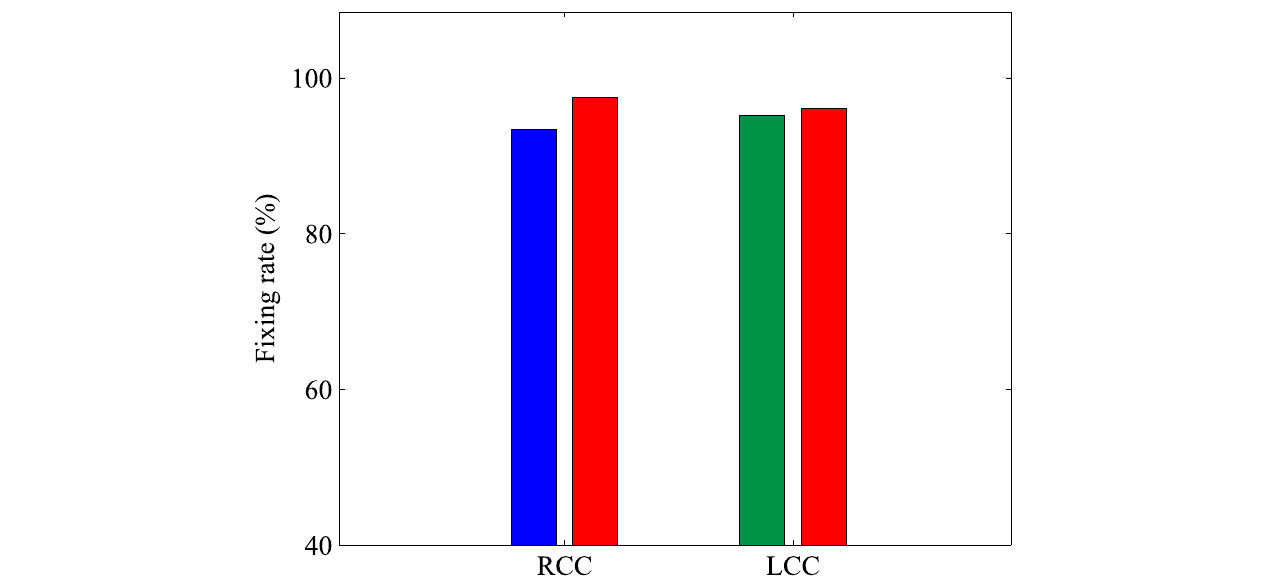}
\caption{Fixing rates of ambiguity resolution for GNSS stations. In AB experiment (RCC), GNSS stations are J001 (red) at JTFS and L002 (blue) at LTFS. In AA experiment (LCC), GNSS stations are J001 (red) and J002 (green) at JTFS.}
\end{figure}


\subsection*{Fixing rates of ambiguity resolution.}
The fixing rate of ambiguity resolution is a useful indicator to evaluate the performance of PPP-AR, expressed as\cite{34}
\renewcommand\theequation{S\arabic{equation}}  
\setcounter{equation}{0}
\begin{equation}
FR=\frac{N_{fixed}}{N_{total}}\times 100\%
\end{equation}
where $N_{fixed}$ denotes the number of both wide lane and narrow lane ambiguities-fixed, $N_{total}$ represents the number of total ambiguities-float. FIGURE. S2 shows the fixing rates of ambiguity resolution in our experiments. The fixing rates of all stations are larger than 93.4\% and the fixing rate of J001 at JTFS in AB experiment can reach 97.6\%. These demonstrate that the carrier phase ambiguities in our experiments have been fixed effectively and the results resolved by PPP-AR are reliable and accurate. 

\bigskip

\subsection*{Transformation from time difference to frequency difference.}
Consider a signal generator whose instantaneous output $V(t)$ can be written as\cite{35}
\begin{equation}
V(t)=\left[V_0+\varepsilon (t)\right]sin\left[2\pi f_0t+\varphi (t)\right]
\end{equation}
where $V_0$ and $f_0$ are the nominal amplitude and frequency, $t$ is the time variable, $\varepsilon (t)$ and $\varphi (t)$ are the instantaneous fluctuations of amplitude and phase, respectively. The instantaneous fractional frequency deviation from the nominal frequency, $y(t)$, is defined as\cite{35}
\begin{equation}
y(t)=\frac{f(t)-f_0}{f_0}=\frac{\dot{\varphi}(t)} {2\pi f_0}
\end{equation}
where $\dot{\varphi}(t)=\frac {d\varphi (t)}{dt}$ is sufficiently small for $t$. And the instantaneous fractional phase deviation is defined as\cite{35}
\begin{equation}
x(t)=\frac{\varphi (t)}{2\pi f_0}
\end{equation}
From equations (S3) and (S4) it reads
\begin{equation}
y(t)=\frac{dx(t)}{dt}
\end{equation}
Combining equations (S2) and (S4), one obtains
\begin{equation}
V(t)=\left[V_0+\varepsilon (t)\right]sin\left[2\pi f_0\left(t+x(t)\right)\right]
\end{equation}
where $x(t)$ presents the reading deviation of a clock running from the signal given by equation (S2), and here we call it time deviation.

Since any frequency-measurement technique does involve a finite time interval over which the measurement is performed, the instantaneous fractional frequency deviation is not an observable. Therefore, the average value of fractional frequency deviation $\bar y_{\tau_k+\tau}$over a time interval $\tau$ beginning at $\tau_k$ provides a more useful quantity directly related to an experimental result, and it can be expressed as\cite{36}
\begin{equation}
\bar y_{\tau_k+\tau}= \frac {1} {\tau} \int^{\tau_k+\tau}_{\tau_k} y(t)dt=\frac {1} {\tau} \int^{\tau_k+\tau}_{\tau_k} dx(t)=\frac {x(\tau_k+\tau)-x(\tau_k)} {\tau}
\end{equation}
where the average value of fractional frequency deviation denotes the change rate of time deviation from $\tau_k$ to $\tau_k+\tau$.

To compare two clocks, we determine the frequency difference between two clocks by subtracting one fractional frequency deviation from another one. In the time interval $\left[0,n\tau \right]$, the average value of frequency difference between two clocks can be expressed as:
\begin{equation}
\bar y^{12}_{dif}= \frac {1} {n} \sum^n_{i=1}\left[\bar y^{2}_{i\tau} - \bar y^{1}_{i\tau}\right] = \frac {{\left[x^2(n\tau)-x^1(n\tau)\right]}-{\left[x^2(0)-x^1(0)\right]}} {n\tau}
\end{equation}

$\,$

$\,$


\bigskip

\begin{thebibliography}{36}
\bibitem{1} Einstein, A. Die Feldgleichungen der Gravitation. Sitzungsberichte der Königlich Preu$\beta$ischen Akad. der Wissenschaften 1, 844–847 (1915).
\bibitem{2} Bjerhammar, A. On a relativistic geodesy. Bull. Géodésique 59, 207–220 (1985).
\bibitem{3} Shen, W., Ning, J., Liu, J., Li, J. \& Chao, D. Determination of the geopotential and orthometric height based on frequency shift equation. Nat. Sci. 03, 388–396 (2011).
\bibitem{4} Takamoto, M. et al. Test of general relativity by a pair of transportable optical lattice clocks. Nat. Photonics 14, 411–415 (2020).
\bibitem{5} Takano, T. et al. Geopotential measurements with synchronously linked optical lattice clocks. Nat. Photonics 10, 662–666 (2016).
\bibitem{6} Lisdat, C. et al. A clock network for geodesy and fundamental science. Nat. Commun. 7, 1–7 (2016).
\bibitem{7} Grotti, J. et al. Geodesy and metrology with a transportable optical clock. Nat. Phys. 14, 437–441 (2018).
\bibitem{8} Hofmann-Wellenhof, B. \& Moritz, H. Physical Geodesy. (Physical Geodesy, by B. Hofmann-Wellenhof and H. Moritz. Approx. 500 p. 100 illus. 3-211-23584-1. Berlin: Springer, 2005., 2005).
\bibitem{9} Campbell, S. L. et al. A Fermi-degenerate three-dimensional optical lattice clock. Science (80-. ). 358, 90–94 (2017).
\bibitem{10} 	McGrew, W. F. et al. Atomic clock performance enabling geodesy below the centimetre level. Nature 564, 87–90 (2018).
\bibitem{11} Oelker, E. et al. Demonstration of 4.8 × 10$^{-17}$ stability at 1 s for two independent optical clocks. Nat. Photonics 13, 714–719 (2019).
\bibitem{12} 	Brewer, S. M. et al. $^{27}$Al$^+$  Quantum-Logic clock with a systematic uncertainty below 10$^{-18}$. Phys. Rev. Lett. 123, 33201 (2019).
\bibitem{13} Bloom, B. J. et al. An optical lattice clock with accuracy and stability at the 10$^{-18}$ level. Nature 506, 71–75 (2014).
\bibitem{14} Huang, Y. et al. Liquid-Nitrogen-Cooled Ca$^+$ optical clock with systematic uncertainty of 3×10$^{-18}$. Phys. Rev. Appl. 17, 1 (2022).
\bibitem{15} Nicholson, T. L. et al. Systematic evaluation of an atomic clock at 2 × 10$^{-18}$ total uncertainty. Nat. Commun. 6, (2015).
\bibitem{16} Bothwell, T. et al. Resolving the gravitational redshift across a millimetre-scale atomic sample. 602, (2022).
\bibitem{17} Allan, D. W. et al. Accuracy of international time and frequency comparisons via global positioning system satellites in common-view. IEEE Trans. Instrum. Meas. 34, 118–125 (1985).
\bibitem{18} 	Fujieda, M. et al. Carrier-phase-based two-way satellite time and frequency transfer. IEEE Trans. Ultrason. Ferroelectr. Freq. Control 59, 2625–2630 (2012).
\bibitem{19} Petit, G. The TAIPPP pilot experiment. 2009 IEEE Int. Freq. Control Symp. Jt. with 22nd Eur. Freq. Time Forum 116–119 (2009) doi:10.1109/FREQ.2009.5168153.
\bibitem{20} Kouba, J. A Guide to using International GNSS Service ( IGS ) products. Geod. Surv. Div. Nat. Resour. Canada Ottawa 6, 34 (2009).
\bibitem{21} 	Geng, J., Chen, X., Pan, Y. \& Zhao, Q. A modified phase clock/bias model to improve PPP ambiguity resolution at Wuhan University. J. Geod. 93, 2053–2067 (2019).
\bibitem{22} 	Laurichesse, D., Mercier, F., Berthias, J. P., Broca, P. \& Cerri, L. Integer ambiguity resolution on undifferenced GPS phase measurements and its application to PPP and satellite precise orbit determination. Navig. J. Inst. Navig. 56, 135–149 (2009).
\bibitem{23} Ge, M., Gendt, G., Rothacher, M., Shi, C. \& Liu, J. Resolution of GPS carrier-phase ambiguities in precise point positioning (PPP) with daily observations. J. Geod. 82, 389–399 (2008).
\bibitem{24} Petit, G. et al. 1 × 10$^{-16}$ frequency transfer by GPS PPP with integer ambiguity resolution. Metrologia 52, 301–309 (2015).
\bibitem{25} 	Petit, G. Sub-10–16 accuracy GNSS frequency transfer with IPPP. GPS Solut. 25, 1–9 (2021).
\bibitem{26} Kopeikin, S. M. et al. Chronometric measurement of orthometric height differences by means of atomic clocks. Gravit. Cosmol. 22, 234–244 (2016).
\bibitem{27} Cheng, P. et al. Measuring height difference using two-way satellite time and frequency transfer. Remote Sens. 14, 1–15 (2022).
\bibitem{28} Su, K. \& Jin, S. Triple-frequency carrier phase precise time and frequency transfer models for BDS-3. GPS Solut. 23, (2019).
\bibitem{29} Geng, J. et al. PRIDE PPP-AR: an open-source software for GPS PPP ambiguity resolution. GPS Solut. 23, 1–10 (2019).
\bibitem{30} Ince, E. S. et al. ICGEM-15 years of successful collection and distribution of global gravitational models, associated services and future plans. Earth System Science Data (2019) doi:10.5194/essd-2019-17.
\bibitem{31} Physics, Q., National, H., Astronomical, X., Standards, F. P. \& Time, N. Free-space dissemination of time and frequency with 10-19 instability over 113 km. (2022) doi:10.1038/s41586-022-05228-5.
\bibitem{32} Petit, G. \& Arias, E. F. Use of IGS products in TAI applications. J. Geod. 83, 327–334 (2009).
\bibitem{33} Landskron, D. \& Böhm, J. VMF3/GPT3: refined discrete and empirical troposphere mapping functions. J. Geod. 92, 349–360 (2018).
\bibitem{34} Pan, L., Xiaohong, Z. \& Fei, G. Ambiguity resolved precise point positioning with GPS and BeiDou. J. Geod. 91, 25–40 (2017).
\bibitem{35} Barnes, J. A. et al. Characterization of frequency stability. IEEE Trans. Instrum. Meas. IM–20, 105–120 (1971).
\bibitem{36} Rutman, J. Characterization of phase and frequency instabilities in precision frequency sources: fifteen years of progress. Proc. IEEE 66, 1048–1075 (1978).
\end{thebibliography}
\end{document}